\documentclass[5p,11pt]{elsarticle}

\usepackage{hyperref,graphicx}
\usepackage{lineno}
\usepackage{subcaption}
\journal{Nuclear Instruments and Methods A}

\begin{document}

\begin{frontmatter}

\title{The Notre-Dame Cube: An active-target time-projection chamber for radioactive beam experiments and detector development}

\author[1]{T. Ahn}
\cortext[mycorrespondingauthor]{Corresponding author}
\ead{tan.ahn@nd.edu}
\author[1]{J. S. Randhawa}
\author[1]{S. Aguilar}
\author[1]{D. Blankstein}
\author[1,2]{L. Delgado}
\author[1]{N. Dixneuf}
\author[1]{S. L. Henderson}
\author[1]{W. Jackson}
\author[1]{L. Jensen}
\author[1]{S. Jin}
\author[1]{J. Koci}
\author[1]{J. J. Kolata}
\author[1]{J. Lai}
\author[1]{J. Levano}
\author[1]{X. Li}
\author[1]{A. Mubarak}
\author[1]{P. D. O'Malley}
\author[1,4]{S. Rameriz Martin}
\author[1,5]{M. Renaud}
\author[1]{M. Z. Serikow}

\author[1,6]{A. Tollefson}
\author[1]{J. Wilson}
\author[1]{L. Yan}
\address[1]{Department of Physics, University of Notre Dame,
225 Nieuwland Science Hall, Notre Dame, IN 46556}
\address[2]{Physics and Astronomy Department, Vassar College, 124 Raymond Avenue, Poughkeepsie, NY 12604}
\address[3]{Physics and Astronomy Department, University of Minnesota Duluth, 1049 University Drive,
Duluth, MN 55812}
\address[4]{Department of Physics, Transylvania University, 300 North Broadway, 
Lexington, KY 40508}
\address[5]{KU Leuven, Instituut voor Kern- en Stralingsfysica, 3001 Leuven, Belgium}
\address[6]{Department of Physics and Engineering, Bethel University,
3900 Bethel Drive
St. Paul, MN 55112}

\begin{abstract}
Active-target detectors have the potential to address the difficulties associated with the low intensities of radioactive beams. We have developed an active-target detector, the Notre Dame Cube (ND-Cube), to perform experiments with radioactive beams produced at \textit{TwinSol} and to aid in the development of active-target techniques.
Various aspects of the ND-Cube and its design were characterized. The ND-Cube was commissioned with a $^{7}$Li beam for measuring $^{40}$Ar+$^{7}$Li fusion reaction cross sections and investigating $^{7}$Li($\alpha$,$\alpha$)$^{7}$Li scattering events.
The ND-Cube will be used to study a range of reactions using light radioactive ions produced at low energy.
\end{abstract}


\end{frontmatter}

%\linenumbers

\section{Introduction}
Radioactive ion beam facilities provide unique opportunities to access the unexplored nuclear landscape \cite{Blumenfeld2013} and nuclear reactions provide one of the most important tools for such studies. Radioactive-ion beams are challenging to produce, with their intensities dropping by orders of magnitude as one moves away from the valley of stability. One of the main aims of state-of-the-art detector development is to obtain a high efficiency and luminosity (large target thickness) for low beam intensities, without sacrificing energy and angular resolution.
Active-target detectors 
address these low beam intensities by increasing overall target thickness and  by providing large angular coverage and low-energy thresholds for detecting particles that would normally be
stopped in a solid target \cite{Beceiro2015}. These properties make active-target detectors  suitable for a large variety of nuclear physics studies with radioactive beams, which include the topics of nuclear structure, nuclear astrophysics, nuclear reactions, and rare-decays \cite{Bazin2020}. There is an active research program investigating these topics at the Institute for Structure and Nuclear Astrophysics (\mbox{ISNAP}) at the University of Notre Dame using light in-flight radioactive beams \cite{Ahn2016}.  These beams are provided by \textit{TwinSol}, a pair of superconducting solenoids that are used as a separator for secondary beams produced in-flight \cite{Becchetti2003}. To take full advantage of the available radioactive beams from \textit{TwinSol} for our science program, we have developed the Notre Dame Cube (ND-Cube) active-target time-projection chamber (TPC). The ND-Cube at \textit{TwinSol} will be used for fusion cross-section measurements, direct measurements of ($\alpha$,$p$) and ($p$,$\alpha$) reactions for nuclear astrophysics, and resonant scattering and transfer reactions for nuclear structure studies. The ND-Cube will also be used for the development of active-target detector techniques. This development includes the testing of new anode pad-plane designs for different Micropattern Gas Detectors (MPGDs) such as the Micromegas \cite{Giomataris1996,Charpak2002} and the Gas Electron Multiplier (GEM) \cite{Sauli1997,Sauli2016}. Future testing for of novel MPGDs such as the TIP-HOLE detector \cite {Randhawa2020} will be considered. In addition to MPGDs, the ND-Cube will be used to optimize front-end electronics and the full electronics chain, test trigger schemes, and benchmark analysis techniques. There are plans for future upgrades of the ND-Cube that include combining the ND-Cube with auxiliary detectors and the use of lasers for drift-velocity calibration. In this paper, we provide a detailed description of the ND-Cube and the first results from an in-beam commissioning, where we measured the cross section of the $^{7}$Li + $^{40}$Ar fusion reaction and we recorded tracks from a $^7$Li($\alpha$,$\alpha$) scattering reaction using inverse kinematics. We also present gas-gain measurements with a double-layer Thick-GEM (THGEM).

\section{Detector Description}

The ND-Cube is an active-target time-projection chamber that has a rectangular geometry. The incoming beam enters into the rectangular volume through a beam window and the electric field is in the vertical direction i.e. perpendicular to the beam axis, resulting in the downward drift of the primary ionization electrons. This is similar to the geometries of other active-target detectors such as MAYA \cite{Demonchy2007}, ACTAR \cite{Mauss2019}, various MUSIC detectors \cite{Carnelli2015}, MSTPC and its variants \cite{Mizoi1999,Hashimoto2006,Yamaguchi2010}, and MAIKO \cite{Furuno2018}. Cylindrical and rectangular geometries for active-target TPCs have both advantages and disadvantages which depend on the details of an experiment. One  advantage of a rectangular geometry is having the drift perpendicular to the beam axis can help alleviate beam-induced space-charge effects which have been  observed when drift direction is along the beam direction \cite{Randhawa2019}. A second advantage is the determination of the position of a particle along the beam axis is not dependent on the determination of the drift velocity. A disadvantage may be the implementation of an internal geometrical trigger is more difficult due to the breaking of cylindrical symmetry around the beam axis.

\begin{figure*}[htbp] %  figure placement: here, top, bottom, or page
   \centering
   \includegraphics[width=7.2in]{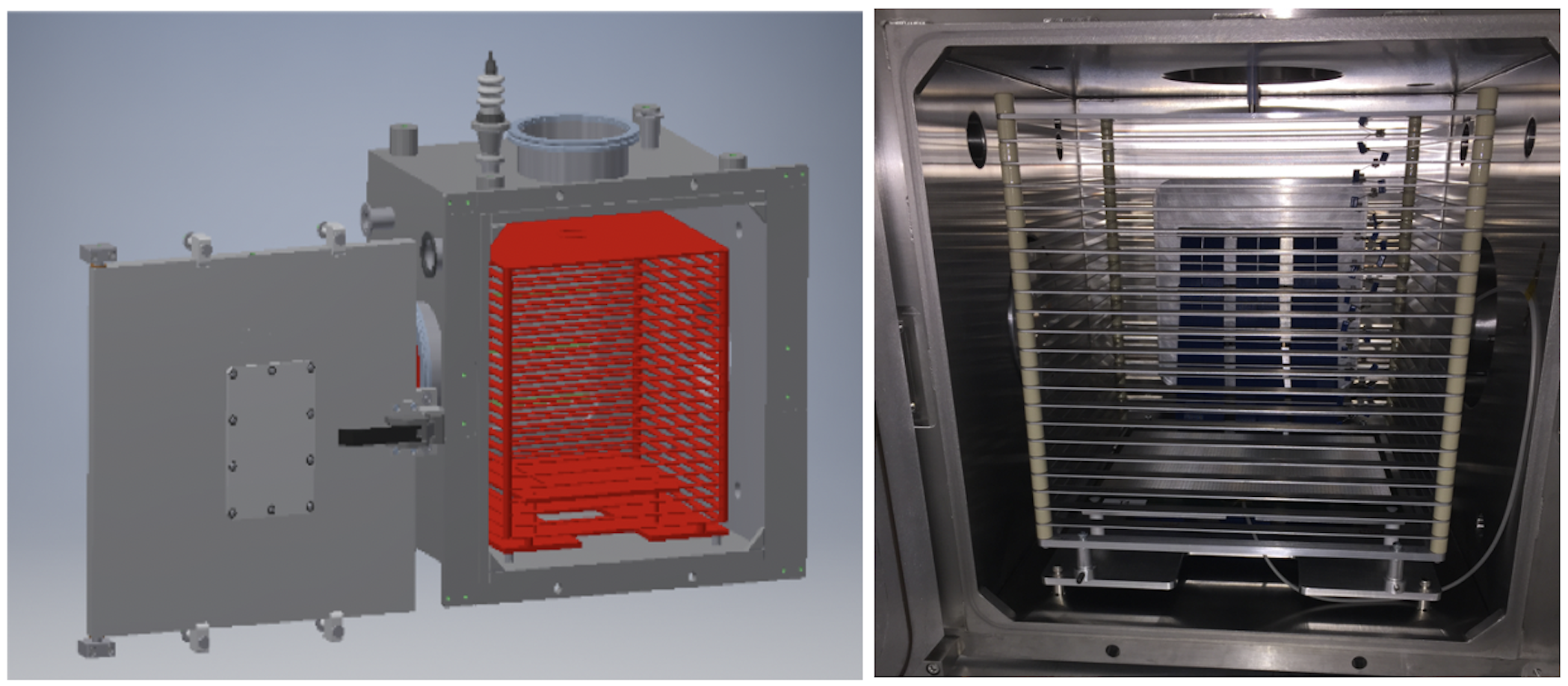} 
   \caption{(Color Online) A mechanical drawing of the ND-Cube with its front door removed is shown on the left. The field cage and Micromegas/Anode pad plane is shown in red. A photograph of the inside of the chamber showing the field cage and Micromegas is shown on the right.}
   \label{fig:cube_drawing}
\end{figure*}

A drawing of the ND-Cube and photograph of the field cage and Micromegas is shown in Fig.~\ref{fig:cube_drawing}. The field cage and anode pad plane are housed in a cubical vacuum chamber that has inner dimensions of 40 cm $\times$ 40 cm $\times$ 40 cm. Two ISO 160 ports, one of which is used as a beam port, are located on the sides of the chamber and an ISO 200 port is located on the top of the chamber. The anode pad plane inside the field cage defines the active area of the detector. The active area is approximately 20 cm $\times$ 30 cm $\times$ 30 cm.  Each major component of the detector will be described in the sections below.

\subsection{Beam Window}

The beam window consists a single-piece cylindrical design with a 2-cm opening that is centered on an ISO 160 port plus a thin foil covering the opening. The beam window was designed to also be a centering ring for the ISO 160 connection. The design protrudes 6 cm downstream to bring the opening of the beam window closer to the active area. A thin foil separates the detector gas from the vacuum of the beamline and various materials can be used for the foil including aluminized mylar and para-aramid of various thicknesses. These materials are epoxied to the ISO 160 piece thus forming the complete beam window. The edges of the opening are rounded in order to keep sharp edges from tearing the thin beam-window material. The diameter of the entrance has been chosen to maximize the transmission of large-emittance radioactive beams to the active area of ND-Cube while allowing for the use of thin window foils. We have successfully used 3-$\mu$m aluminized mylar with pressures up to 250 Torr and 12-$\mu$ para-aramid up to 1 atm.

\subsection{Field Cage}

The field cage provides the uniform electric field needed for the transport of ionization electrons to the anode pad plane. 
The field cage is composed of a solid square cathode plate and 21 square aluminum rings that are held in place with 1.27 cm (0.5 inch) tall cylindrical ceramic 
standoffs. The total height of the field cage from cathode to the first layer of the MPGD is 33.3 cm, which defines the height of the active volume. A custom high-voltage feedthrough is placed in the top ISO 200 port and provides  high voltage to the top plate through a spring loaded electrode. Each ring of the field cage is connected by a 20-M$\Omega$ resistor to form a resistor chain. The total resistance of the resistor chain is 440 M$\Omega$. For typical cathode voltages from 3.0-7.5 kV, this amounts to a current of 6.8-17.0 $\mu$A and a total power dissipation of 0.02-0.13 W. The top surface of the MPGD
is placed at the same height as the second-to-last ring in the chain. This ring is biased to the same voltage as the top of the MPGD and thus ensures a uniform electric field down from the cathode to the MPGD used for the amplification of the ionization electrons. 

% Temp measurements need to be done at some point
%The temperature of the gas was also monitored using a thermocouple gauge and found to fluctuate on the order of xx C$^{\circ}$ over a 24-hour period.

\begin{figure}[htbp] %  figure placement: here, top, bottom, or page
   \centering
   \includegraphics[width=3.5in]{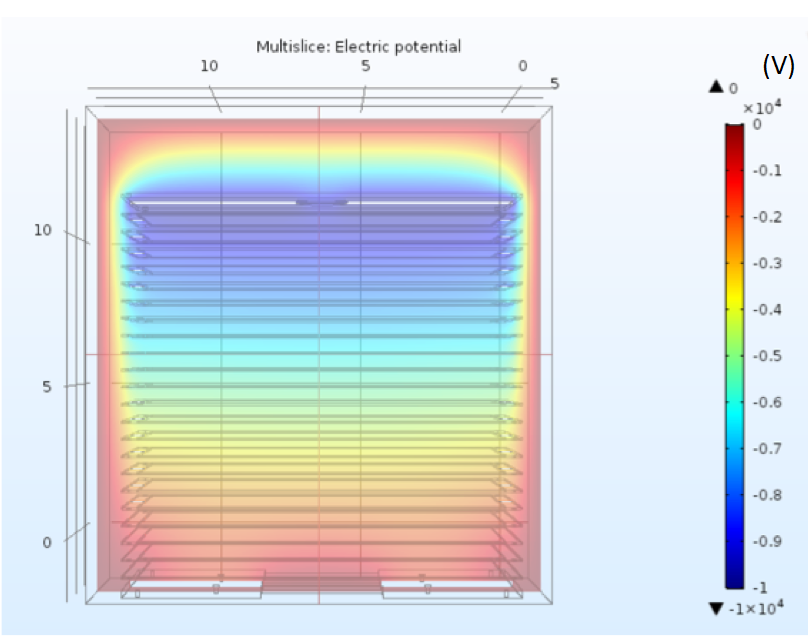} 
   \caption{(Color Online) The calculated electric potential of the field cage using finite-element analysis program COMSOL.}
   \label{fig:field_cage_potential}
\end{figure}

The electric potential and electric field of the field cage was simulated using the finite-element analysis program COMSOL \cite{comsol}, to
look for regions of high electric field and to use the electric fields for electron drift simulations. A plot of the electric potential in the active volume and the region outside the field cage is shown in Figure \ref{fig:field_cage_potential}. The simulated electric field appears to be very uniform for the active region. 

We have tested the maximum achievable electric field strength for a given gas and gas pressure. 
For this test, we used 500 Torr of He:CO$_2$(90:10) gas, which gives a drift field of 287 V/cm (0.57 V/cm/Torr) with a cathode voltage of $-9.5$ kV . The higher cathode voltages for a given pressure were limited by sparking of the gas. The region of sparking needs to be further investigated, but  the field cage simulations shows that the region the rings closest to the cathode and the chamber wall show the highest field strengths. In future iterations of the field cage, we will consider increasing this distance to prevent discharging of the cathode to the chamber walls which will allow us to achieve higher drift fields.

\subsection{Anode Pad-Plane Design and Primary Electron Amplification}

\begin{figure}[htbp] %  figure placement: here, top, bottom, or page
   \centering
   \includegraphics[width=2.5in, angle=270]{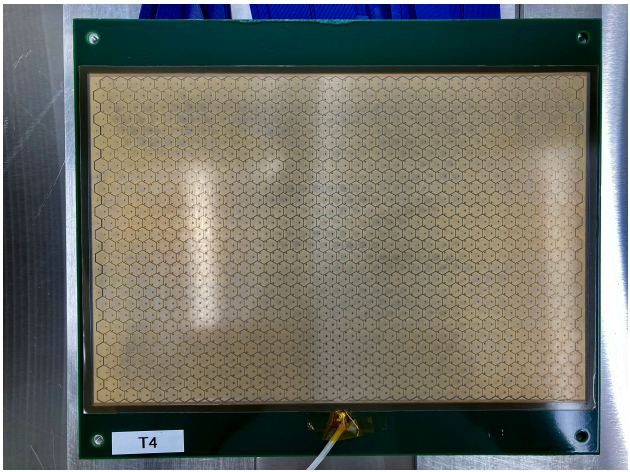} 
   \caption{(Color Online) A photograph of the Micromegas micro-pattern gas detector. The 1008 hexagonal anode pad-plane design can be seen.}
   \label{fig:micromegas}
\end{figure}

The ND-Cube's anode pad plane is a printed-circuit board with conducting pads on the upper surface. The design consists of a tiling of 1008 hexagons. Each hexagonal pad has a 4 mm side and is 7 mm across. A picture of the pad-plane tiling as part of our Micromegas is shown
in Fig.~\ref{fig:micromegas}. The pads were finished with an immersion gold process to prevent the oxidation of the copper layer.  
The dimensions of the total active area are 230 mm $\times$ 275 mm. The number of hexagons were chosen to have a fine granularity and yet maximize the active volume given the total number of electronic channels that were available. The pads were routed to a total of nine fine-pitch connectors on the bottom surface. The anode pad plane connects to the feed-through board, ZAP A, through 120-pin cables where the signal is then routed to the front-end electronics (see Sec.~\ref{sec:zap}).

\begin{figure}[htbp] %  figure placement: here, top, bottom, or page
  \centering
  \includegraphics[width=2.5in, angle=0]{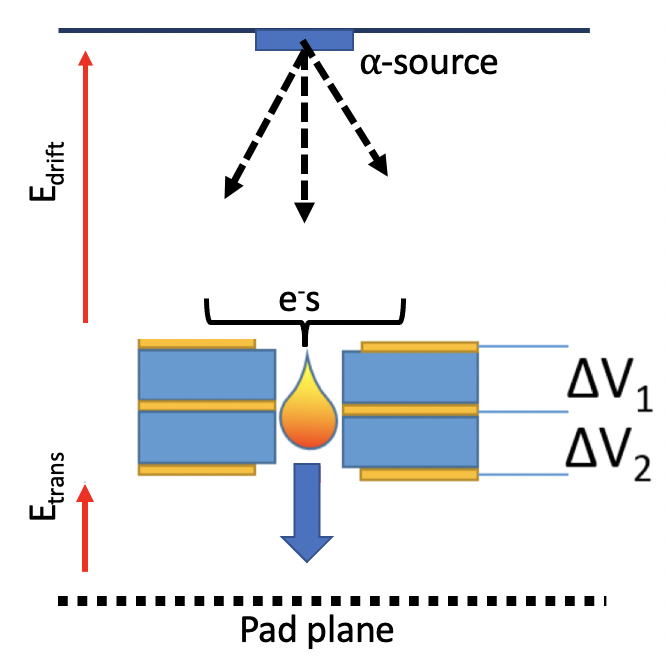} 
  \caption{(Color Online) A schematic drawing for the double-layer THGEM. The $\alpha$-source setup is shown used for gas-gain measurements is shown. The drift field $E_{\rm drift}$ and transfer field to the anode pad plane $E_{trans}$ are shown (not to scale).}
  \label{fig:thgem}
\end{figure}

\begin{figure}[htbp] %  figure placement: here, top, bottom, or page
  \centering
  \includegraphics[width=3.5in]{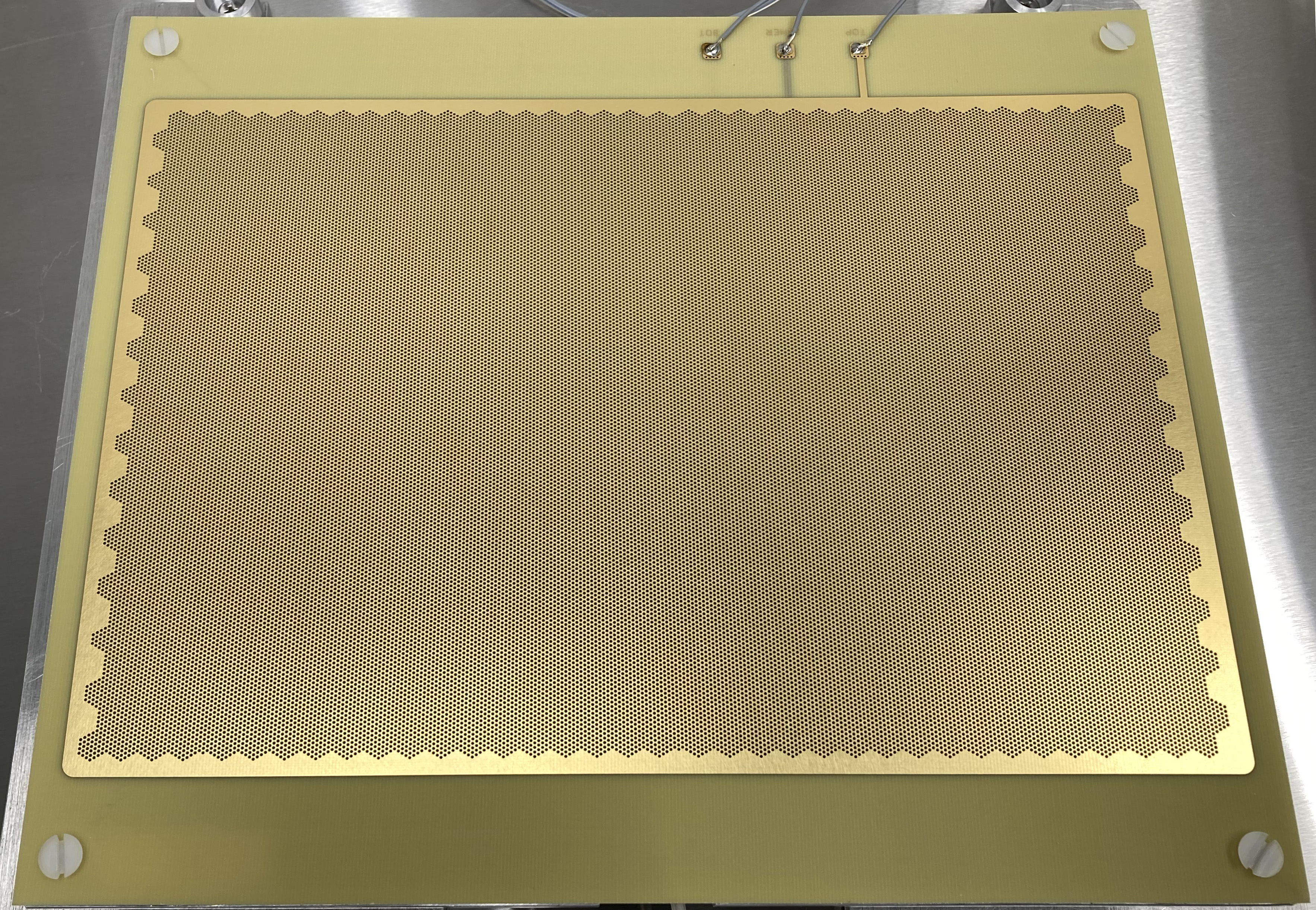} 
  \caption{(Color Online) A photograph of the double-layer THGEM. The active region is corresponds to the overall shape of the pads of the anode pad plane.}
  \label{fig:thgem_photo}
\end{figure}

We use two different MPGDs for the ND-Cube. The first is a Micromegas and the second is a double-layer thick GEM (THGEM). Our Micromegas was produced by CEA Saclay and used the anode pad plane. A micromesh is held in place above the pads with Kapton pillars that keep a uniform 128 $\mu$m gap, which forms the amplification region. The Kapton pillars holding the mesh can be seen as the small dots that cover the pad plane (Fig.~\ref{fig:micromegas}). The double-layer THGEM, which was produced at CERN, can be used with a bare anode pad plane or in combination with a Micromegas to give a dual-stage amplication. A drawing of the double-layer THGEM is shown in Fig.~\ref{fig:thgem}. The THGEM has three layers of copper conductor separated by insulator. The active region of the THGEM was matched to the shape of the pads of the anode pad plane. A photograph of our THGEM is shown in Fig.~\ref{fig:thgem_photo}. The primary electron amplification occurs in the regions between the conducting layers. The total thickness of the THGEM is 1.2 mm, the holes are 0.5 mm in diameter with a 0.1 mm rim and 1.0 mm pitch. The outer conductor was given a Ni/Au finish. The THGEM is typically mounted 1-2 mm above the Micromegas or bare anode pad plane.
% Figure for ThGEM

\subsection{Electronics Chain}

\begin{figure*}[htbp] %  figure placement: here, top, bottom, or page
   \centering
   \includegraphics[width=7.2in]{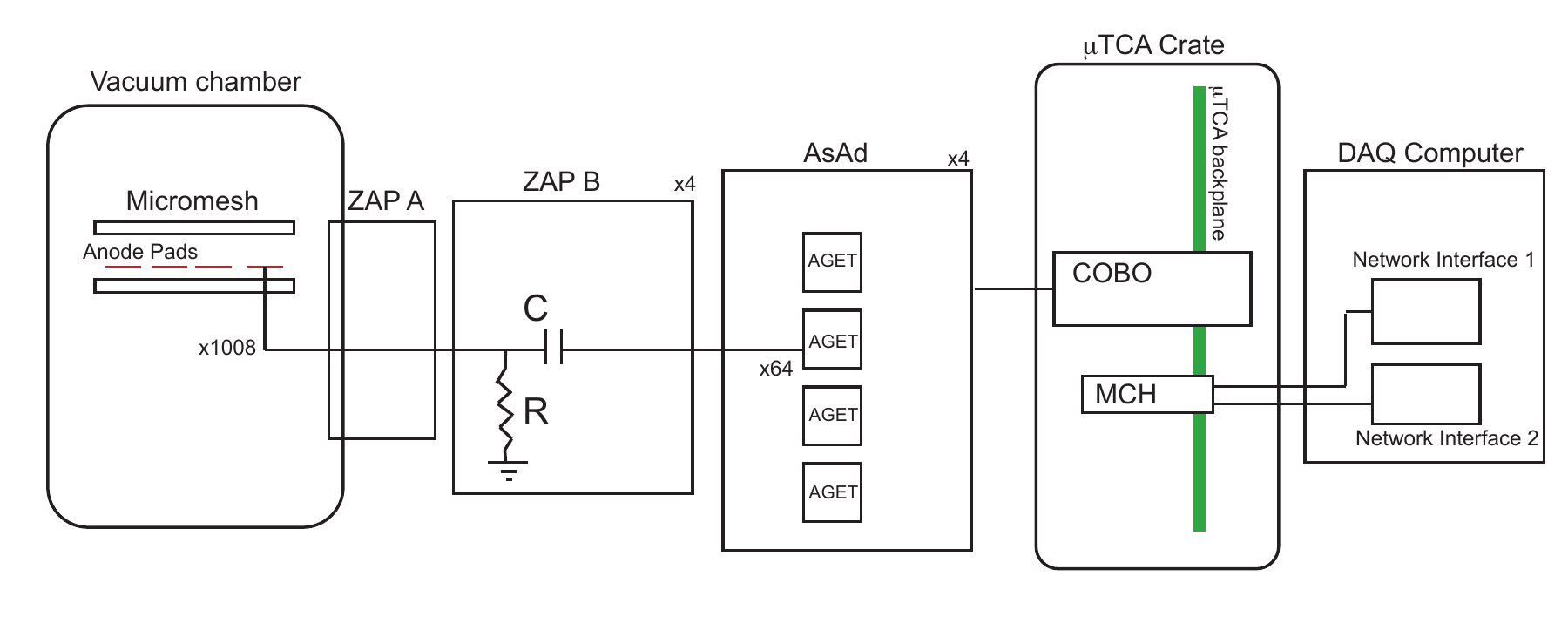} 
   \caption{(Color Online) A schematic diagram of the electronics chain of the ND-Cube from the Micromegas to the data acquisition computer.}
   \label{fig:electronics}
\end{figure*}

Once the primary electrons are amplified, the signal produced in the MPGD is sent through an electronics chain processes and records the signal. A schematic of the electronics chain is shown in Figure \ref{fig:electronics}. The processing and digitization of the signal is done by the General Electronics for TPCs (GET), which is based on Application Specific Integrated Circuit (ASIC) technology developed specifically for TPC experiments with a large number of channels \cite{Pollacco2018}.

\subsubsection{Front-end feedthrough: ZAP boards\label{sec:zap}}

The signals produced by the MPGD inside the ND-Cube's vacuum chamber need to be routed to the front-end electronics that are outside the detector. In order to do this for the signals from the 1008 pads, a two-part feedthrough connection based on printed-circuit boards (PCBs) is used. The first PCB is called the ZAP A and acts as a feedthrough that connects the pad signals to connectors on the outside that match the front-end electronics. The ZAP A also acts as a sealing surface to openings in the vacuum chamber for the inner connectors and uses o-ring seals. The Micromegas or anode pad plane connects to the ZAP A through nine 120-channel ribbon cables. The outside of the ZAP A connects directly to the ZAP B board through 16 72-pin connectors that are the same type as the front-end card, the ASIC and ADC (AsAd) board. The ZAP B has a resistor and capacitor circuit as shown in Fig.~\ref{fig:electronics} that capacitively couples the signal to the input of the AsAd card and and which grounds the anode pads through the resistor. This circuit uses a 220 pF capacitor and 100 M$\Omega$ resistor. Future iterations of the ZAP B will allow for implementing a high-voltage bias on the pads for a larger range of potential differences between the micromesh or THGEM and the anode pad plane. This will enable a more flexible MPGD amplification setup. 

\subsubsection{Front-End Electronics and Data Acquisition}

The signals from the ZAP boards are connected to the AsAd where each channel is connected to one of the four onboard ASIC chips, the ASIC for GET (AGET). Each channel on the AGET has a charge sensitive preamplifier with four gain settings, a shaper and amplifier, pole-zero correction, discriminators for the shaped signal, and a switched capacitor array for digitization of the signal. The signals can be digitized at rates up to 100 MHz with 12-bit resolution. A total of four AsAd cards having a capacity of 1024 channels are used for the 1008 pad signals. Four AsAd cards are connected to one Concentration Board (CoBo). The CoBo can control AGET parameters, implement trigger conditions, and transfer data to the data acquisition computer through the $\mu$TCA crate. Communications and data transfer are through the $\mu$TCA backplane and the interface to the data acquisition computer is through a memory controller (MicroTCA Carrier Hub (MCH)), which uses a Gigabit ethernet connection. Control and implementation of the trigger can be performed in CoBo stand-alone mode or with the MUTANT master trigger logic card. More details can be found in Ref.~\cite{Pollacco2018}.

For an experiment, typically one of two different trigger conditions are used. The first is an externally generated trigger, such as a signal from the micromesh or THGEM, or a multiplicity trigger that is generated internally from the GET electronics. The multiplicity trigger is based on the total number of discriminator signals from individual pads in a chosen time window. Combinations of external and multiplicity triggers can be used. These can be useful in situations where the ND-Cube is used with auxiliary detectors.

\section{Vertex  Resolution and Energy Resolution}

One of the main advantages of active-target detectors is the ability to simultaneously track the beam and reaction products. Event-by-event tracking of particles allows for the determination of the reaction vertex. In the case of active-target detectors where the pad-plane is perpendicular to the beam axis, the vertex resolution depends on how well the time of a charge signal can be converted into a position. This calibration depends on how well the the electron drift velocity is known throughout the detector during the experiment. In the case of the ND-Cube, the pad plane is parallel to the beam axis. Since the pad sizes and distances from the entrance window are known, this allows for the determination of the absolute position of the tracks along the ideal beam axis. The vertex resolution will be on the order of the size of a pad, but with the aid of tracking algorithms, vertex resolutions that are better than the size of a pad can be achieved.
%With the aid of tracking algorithms, even though vertex resolution can be much better, the upper limit on vertex resolution in our case will be equal to pad size.

\begin{figure}
    \centering
    \includegraphics[width=3.7in]{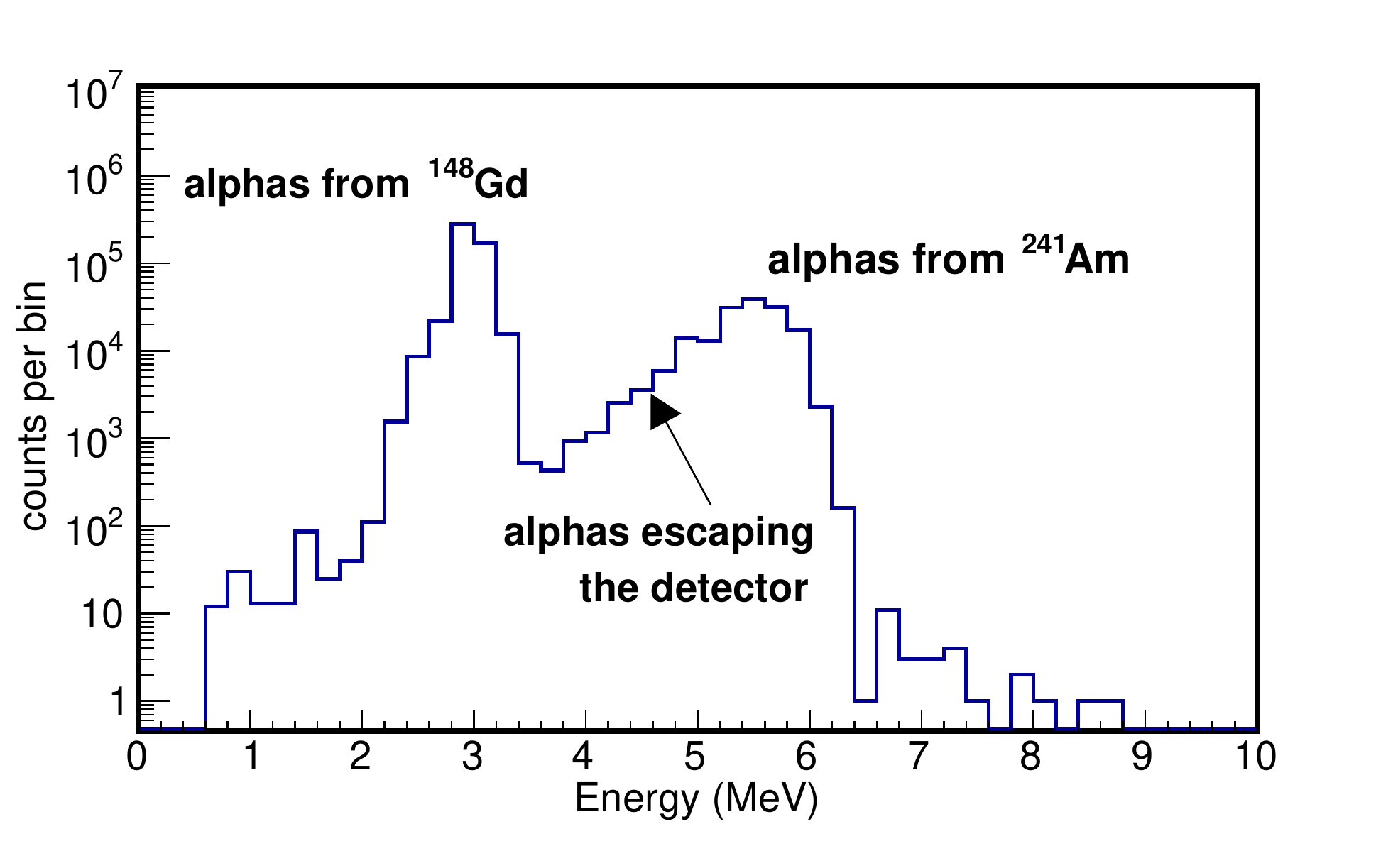}
    \caption{(Color Online) Energy spectrum of $\alpha$-particles from a mixed $\alpha$-source i.e. $^{148}$Gd and $^{241}$Am.}
    \label{fig:alphas}
\end{figure}

Another important measure of detector performance is the total charge resolution of a track. This translates directly to the resolution of the energy deposited by a charged particle in the active region of our detector.
We have performed measurements to determine the energy resolution from total charge integration of an $\alpha$-particle track using a mixed $^{148}$Gd and $^{241}$Am $\alpha$-source. The source had an activity of 2000 Bq. We placed the source above the Micromegas at a height of 20 cm. The source was not collimated and faced the direction of the Micromegas. 
The detector was filled  with P10 gas at a pressure of 400 Torr. At this pressure, the lower energy $\alpha$-particle ($E=3.2$ MeV) from $^{148}$Gd were completely stopped inside the active area. However, depending on the direction of the $\alpha$ particle, many $\alpha$-particles from $^{241}$Am escaped the active region and deposited only part of their energy. The $\alpha$-particle energy spectrum is shown in Figure~\ref{fig:alphas}. Using the known energy of the lower-energy peak at 3.2 MeV, the energy resolution was determined to be 9\% FWHM at 3.2 MeV.

\section{In-Beam Detector Commissioning Experiment}

To test and demonstrate the performance of the ND-Cube under experimental conditions, the ND-Cube was commissioned with a $^7$Li beam. The aim of this test was to demonstrate the use and the readiness of ND-Cube for future fusion studies, as well as scattering studies, which have been demonstrated with other active-target detectors \cite{Kolata2016, Suzuki2012}. Here we report the measurement of $^{7}$Li$+ ^{40}$Ar fusion cross sections using P10 (Ar:CH$_{4}$ (90:10)) gas as a target and $^{7}$Li$(\alpha,\alpha)^{7}$Li using He:CO$_{2}$(90:10) gas.
The $^7$Li beam was produced with the FN Van de Graaff accelerator at the Nuclear Science Laboratory at the University of Notre Dame. A $\Delta E$-$E$ Si telescope was mounted upstream of the ND-Cube entrance window to independently measure the beam properties. To reduce the $^7$Li beam rate in the ND-Cube, two attenuators were used to reduce the beam intensity to 1000-3000 pps. A Micromegas was used for the electron amplification.

\subsection{Measurement of $^7$Li $+$ $^{40}$Ar fusion cross sections \label{sec:fusion}}

\begin{figure*}
    \centering
    \includegraphics[width=7.2in]{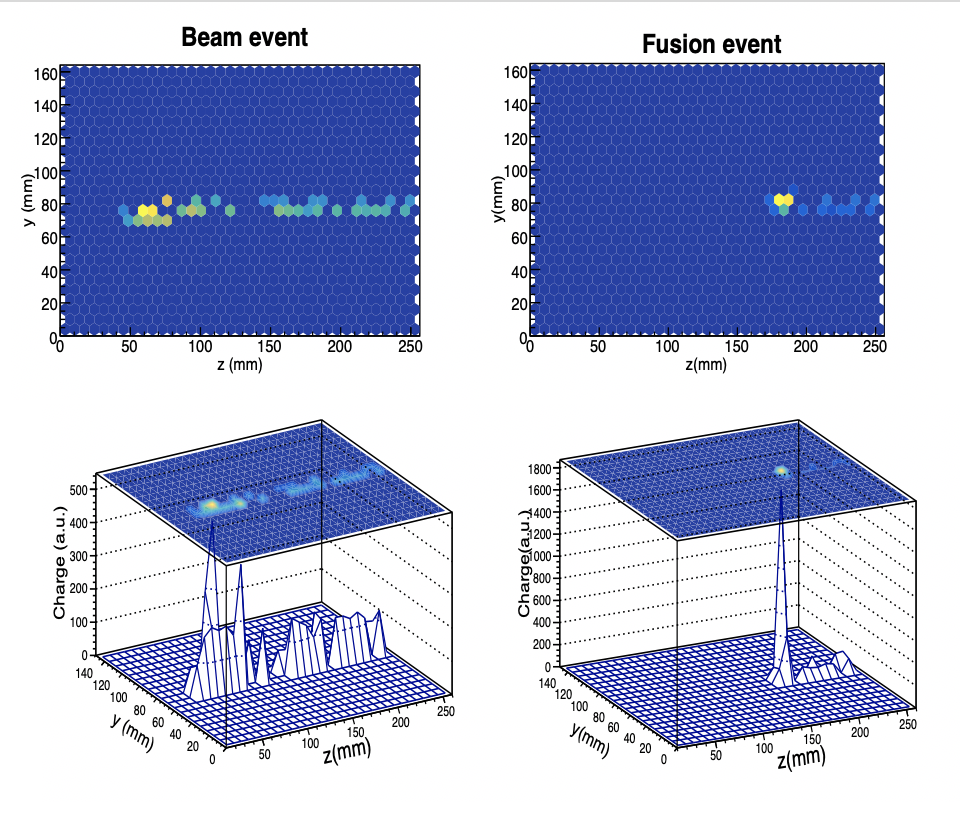}
    \caption{(Color Online) A comparison between a beam and fusion event is shown. The two left plots show an example of a beam event that has a larger range and smaller peak charge. The two right plots show a typical fusion event with a shorter range and much larger energy loss at the position of the fusion event leading to higher peak charge.}
    \label{fig:fusion_beam_tracks}
\end{figure*}

To measure the $^7$Li $+$ $^{40}$Ar fusion cross section, the ND-Cube was filled with 320 Torr of  P10 gas. The field-cage cathode voltage was -1.8 kV which corresponds to a drift field of 44 V/cm.  A $^{7}$Li beam with 21 MeV energy entered through the ND-Cube entrance window. The beam energy and detector pressure were chosen so the beam stops inside the detector close to the end of the active region. When a fusion event occurs, the fusion residue is highly ionizing and has a very short range. This appears as an abrupt and sharp peak in the energy loss signal and is the signature of a fusion event. The top plots in Fig.~\ref{fig:fusion_beam_tracks} show the 2-D projection of a beam track (left) and fusion event(right) onto the anode pad plane with the beam entering from the right. A small number of pad signals are not present due to missing connections the front-end electronics, but the tracks can clearly be seen. The fusion event has a shorter track length as expected. The lower plots in Fig.~\ref{fig:fusion_beam_tracks} show the charge signal for a beam track and fusion event, where this charge signal is proportional to the energy loss of the beam and fusion residue. The second signature of the fusion event, the much larger energy loss of the fusion residue, can be seen in the magnitude of the charge peak.

In order to generate a simple and robust trigger for fusion events, we used the mesh signal from the Micromegas. This signal can be considered to be proportional to the sum of all pads as a function of time. Thus the integrated signal is proportional to the energy deposited by a track in the active region. For unreacted $^7$Li beam particles, this is a well defined energy and thus results in a well defined charge integration pulse height. Fusion events, due to being a completely inelastic collision will result in a smaller Micromegas pulse height. The trigger for fusion events used a single-channel analyzer to accept events for pulse heights that were smaller than those corresponding to the beam.

\begin{figure}
    \centering
    \includegraphics[width=\linewidth]{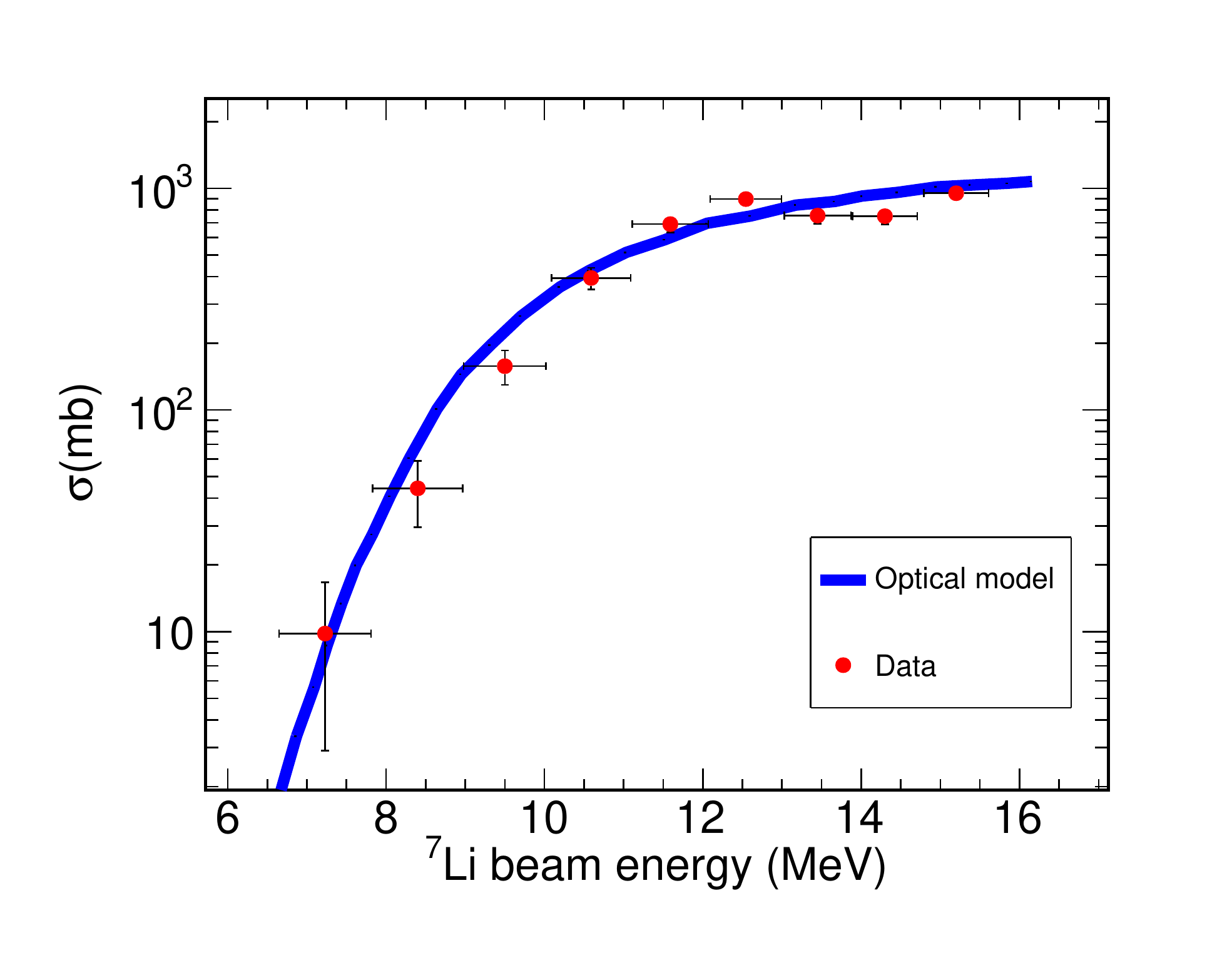}
    \caption{(Color Online) Comparison of measured $^{7}$Li+$^{40}$Ar fusion cross-sections to optical model calculation. Vertical error bars represent statistical uncertainty and horizontal error bars represent the range in energy for a 1.5 cm bin size over which counts are integrated.}
    \label{fig:fusion_cross_section}
\end{figure}

In the offline data analysis, fusion events were identified based on a shorter track length and higher peak charge  as shown in Fig.~\ref{fig:fusion_beam_tracks}. To get the absolute normalization for extracting cross sections, the beam rate was measured using a down-scaled mesh signal. Since the beam was stopped inside the detector, this allowed us to convert the beam track length into beam energy. The pad plane along the beam axis was divided into 1.5 cm bins. Each 1.5 cm bin corresponds to the same number of target atoms, but the energy loss of the beam is different in each bin since it loses more energy the deeper it goes into the detector. The number of fusion events per 1.5 cm bin was obtained and the corresponding fusion cross section is shown in Fig.~\ref{fig:fusion_cross_section}. Here, horizontal error bars represent the range of the beam energy in each 1.5 cm bin. The experimental cross sections were compared to optical-model calculations. The global optical-model parameters were taken from  Ref.~\cite{Cook1982}, which contains data for $^{7}$Li on a wide range of targets for energies in the range of 13-153 MeV. The measured cross-sections show a good agreement within a reasonable range of energy with optical-model calculations (Fig.~\ref{fig:fusion_cross_section}).
 
\subsection{$^7$Li + $\alpha$ elastic scattering}

Using the same 21 MeV $^7$Li beam, we have recorded events from the $^7$Li($\alpha$,$\alpha$)$^7$Li reaction using inverse kinematics. This study was aimed at investigating the effectiveness of the various trigger schemes. 
The ND-Cube was filled with a He:CO$_{2}$ (90:10) gas mixture at a pressure of 400 Torr. The field-cage cathode high voltage was -9.00 kV, which gave a drift field of 256 V/cm. The drift velocity was calculated using the program Magboltz \cite{Biagi1989,Biagi1999} and found to be 1.1 cm/$\mu$s. This drift velocity corresponds to a total drift time from the cathode to the Micromegas of 36.6 $\mu$s. This should allow for a beam rate on the order of $2\times 10^5$ pps but we limited our maximum beam rate to about 5000 pps due to the possible limitations of the data rate in our data acquisition system.

\begin{figure}
\centering
\includegraphics[width=3.5in]{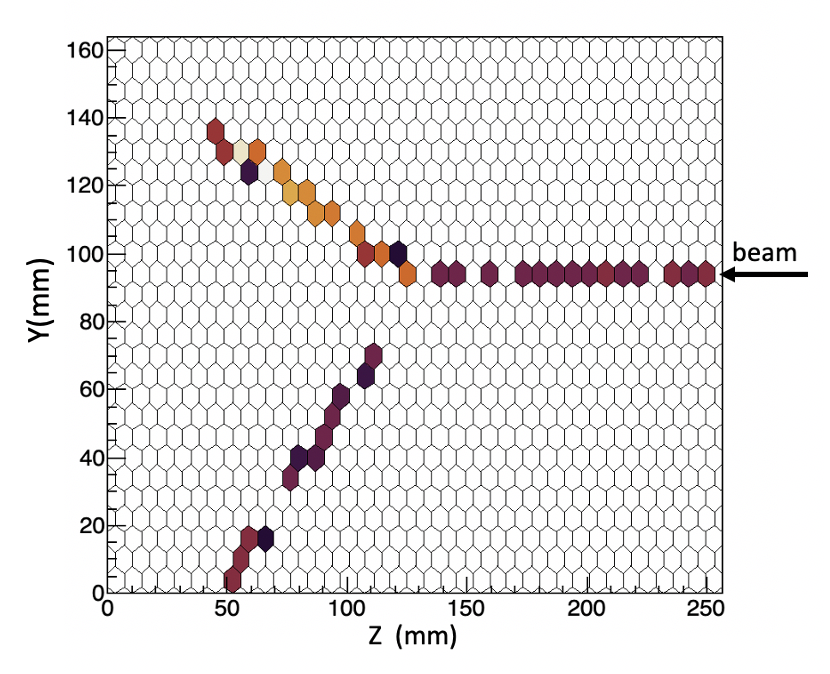}
\caption{(Color Online) 2D-projection of a sample scattering event for the $^{7}$Li +$\alpha$ system. The beam path as well both scattered particle tracks are visible.}
\label{fig:scattering}
\end{figure}

\begin{figure*}
\centering
\includegraphics[width=7.2in]{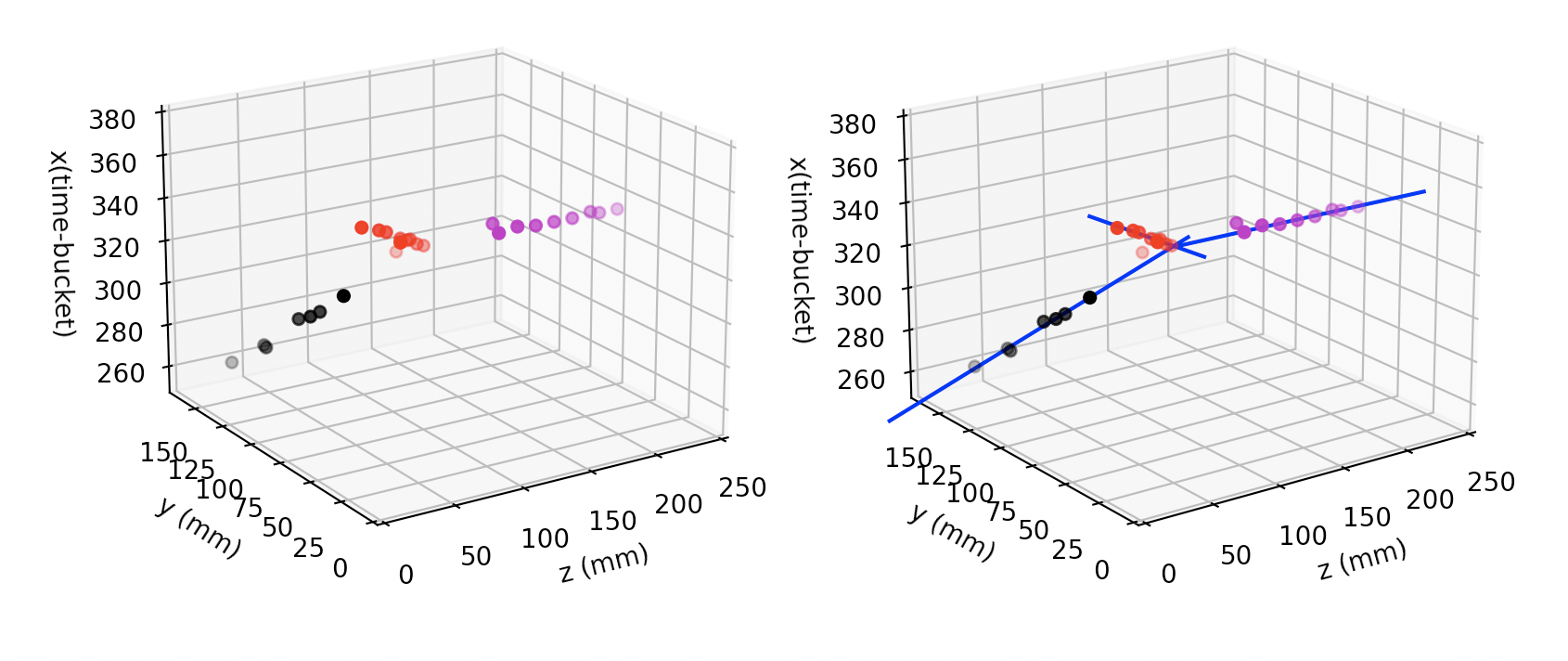}
\caption{(Color Online) This figure shows the implementation of tracking algorithms. Left panel shows implementation of hierarchical clustering where three identified clusters are shown in three different colors. Right panel shows the RANSAC fit to the each identified cluster.}
\label{fig:tracking_algorithm}
\end{figure*}

The trigger condition was based on a multiplicity threshold that corresponds to the number of pads hit in a certain time window. One example of a scattering event is shown in Fig.~\ref{fig:scattering}. To track and identify the particles from such images, we are developing tracking algorithms based on the Python-based hierarchical-clustering algorithm \cite{Pedregosa2011} and the RANSAC algorithm \cite{Ayyad2018}. These algorithms have been successfully used to track the particles in 3D. We used the Agglomerative Clustering algorithm for which the hierarchical clustering uses a bottom up approach.  Each observation starts with its own cluster and clusters are successively merged together. The algorithm merges a pair of clusters that minimally increase a given linkage distance. One can either provide the linkage distance threshold to the algorithm so that clusters outside this threshold are not merged or one can define the number of clusters to be searched for. This allows the identification of different clusters where each cluster corresponds to a distinct particle track. Once different tracks (each as a cluster) are identified, the data points of each track are passed to the RANSAC algorithm. RANSAC provides a robust way to fit a line to the track positions where the algorithm iteratively estimates the line's parameters. At each iteration, RANSAC estimates a model on a random subset and checks whether the estimated model is valid. It then classifies all the data points as either inliers or outliers by calculating their residuals using the estimated model. All data points with residuals smaller than the predefined residual-threshold are considered inliers. Once the best-fit line parameters are determined, the direction and range of a track can be deduced. Fig.~\ref{fig:tracking_algorithm}(a) shows a sample event where hierarchical-clustering identified three different tracks, i.e. the incoming beam track, the scattered beam track and the scattered $^4$He track. Fig.~\ref{fig:tracking_algorithm}(b) shows the implementation of RANSAC for each of the identified clusters. RANSAC fit three different lines, one for each track and also rejected outlier points which were initially identified as a part of a track cluster. Data analysis codes using these algorithms are currently under development and will provide a robust way to track particles and identify reaction channels in future experiments.

\section{Current and Future Developments}

The ND-Cube will serve as a development platform for the use of novel gas mixtures, different types of MPGDs, the use of ancillary detectors, and analysis techniques.
One of the important features of active-target detectors has been the use of pure hydrogen, deuterium, and noble gases as a target and tracking medium \cite{Cortesi2018}. The use of pure gases, such as hydrogen and helium, for scattering experiments and noble gases, such as Ne and Ar, for fusion experiments is advantageous as the data will be free from any undesired background events originating from reactions on the nuclei of a quench gas. The feasibility of using pure gases in active-target TPCs has been demonstrated by the combined use of THGEM layers and a Micromegas \cite{Cortesi2018}. The ND-Cube will be used  to investigate the use of these pure gases and other potentially useful gas mixtures.

\begin{figure}[ht]
\centering
\includegraphics[width=\linewidth]{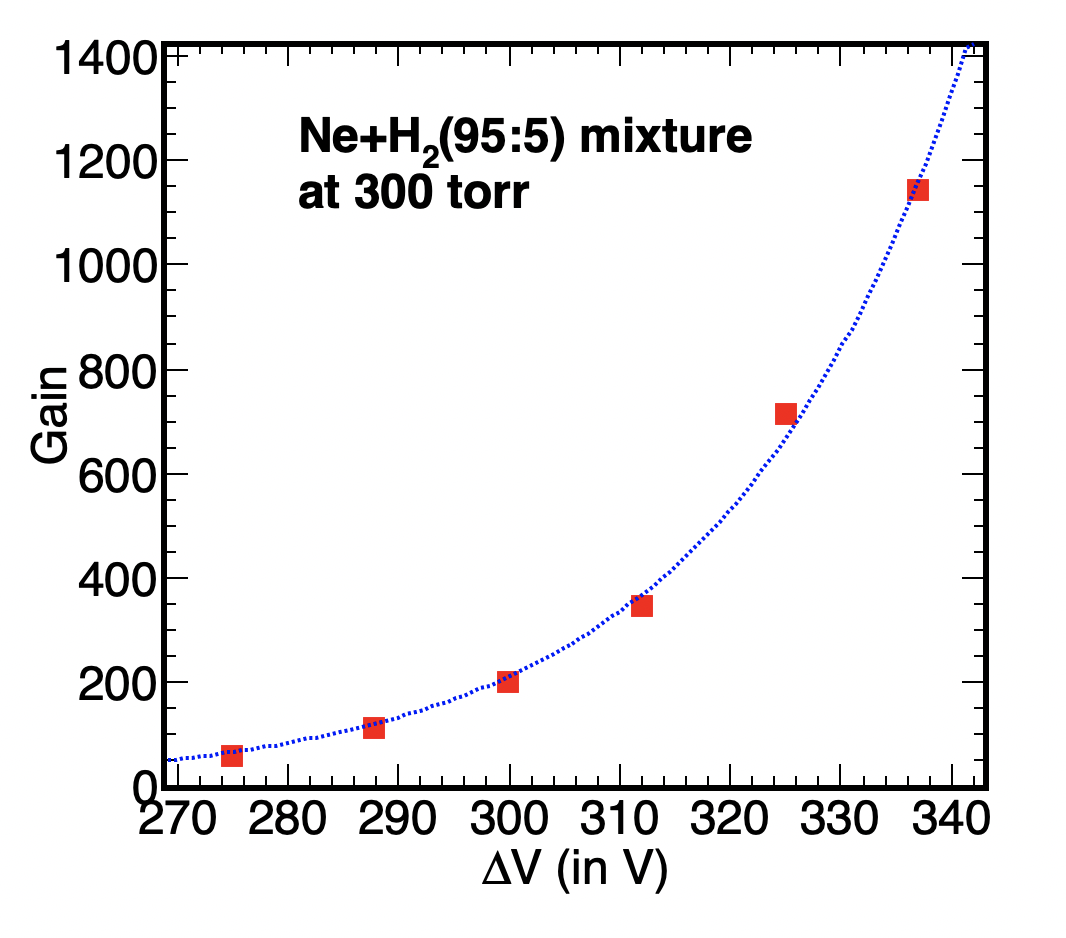}  
\caption{(Color Online) Plot shows the gain curve in 300 torr of Ne:H$_{2}$(95:5) gas mixture. THGEMs were operated in the symmetric mode i.e. voltage difference across the two amplification regions was was same ($\Delta$V=$\Delta$V$_{1}$=$\Delta$V$_{2}$). }
\label{fig:fig_last}
\end{figure}

As a first step in this direction, we provide the first gain measurements using a double-layer THGEM in the ND-Cube using a Ne:H$_{2}$ (95:5) gas mixture. Use of Ne gas as a target for  fusion reactions has been of high interest for astrophysics \cite{Avila2016}. In Ne:H$_{2}$ mixtures, the H$_{2}$ acts as a quench and Penning gas \cite{Buzulutskov2005}  but has very little impact on measurement of fusion reactions using the technique described in Section \ref{sec:fusion}. For the gain measurements, we operated our THGEM in symmetric mode, i.e. the potential difference across the THGEM's two layers was kept same ($\Delta$V$_{1}$=$\Delta$V$_{2}$ in Fig.~\ref{fig:thgem}). The gain curve for a 300-Torr Ne:H$_{2}$(95:5) gas mixture is shown in Fig.~\ref{fig:fig_last}. In future studies, the THGEM in combination with the Micromegas will be used to perform reaction studies with pure gas targets such as He and H$_{2}$. 

Other important future advancements  will be  development and integration of ancillary detectors with the ND-Cube. In many experimental scenarios, light energetic particles will leave the active region of the ND-Cube where particle identification and/or the measurement of total energy may be difficult. The addition of ancillary detectors around the active region can give information, such as total energy, on the particles leaving the active region of the detector. Such ancillary detectors may include silicon detectors, scintillator detectors, or gas detectors. Future developments will include testing designs and integration of such ancillary detectors with the ND-Cube.

\section{Summary and Outlook}

We have developed a new active-target detector, the ND-Cube, and have tested and commissioned the detector for use in nuclear physics experiments at the Nuclear Science Laboratory at the University of Notre Dame. This included measurements with an $\alpha$ source, an in-beam measurement with a $^7$Li beam, and gas gain measurements with a double-layer THGEM. The in-beam $^7$Li experiment was performed using two different target gases for two different studies: a fusion cross section measurement and the recording of two-body $\alpha$ scattering. We measured the $^{7}$Li$+^{40}$Ar fusion cross section using P10 gas, and elastic scattering, $^{7}$Li($\alpha,\alpha$)$^{7}$Li, was observed using He:CO$_{2}$ gas as a target. The $^{7}$Li$+^{40}$Ar fusion cross sections measured in this work are in good agreement with optical model calculations, which provides confidence for the future use of ND-Cube in the study of fusion reactions with the radioactive isotope beams. For the $\alpha$ scattering data, the initial performance of the tracking analysis, currently under development, is shown. The first measurement of the gain of a double-layer THGEM was made for a Ne:H$_2$ mixture, the results of which will be used for future fusion studies with Ne isotopes. Future developments will include use of ancillary detectors. The ND-Cube provides a promising platform to perform nuclear reaction experiments with the radioactive beams from the \textit{TwinSol} facility at Notre Dame. Use of an active-target detector will enhance the capabilities to perform direct reaction measurements such as ($p$,$\alpha$), ($\alpha$,$p$), and ($\alpha$,$n$) reactions, other transfer reactions, and fusion studies. 

\section*{Acknowledgement}

We would like to acknowledge S. Aune at CEA Saclay for the production of our Micromegas detector and R. De Oliveira at CERN for the production of our THGEM.
This work was supported by the National Science Foundation Grant Nos. 1713857 and 2011890, and the University of Notre Dame.

%\section*{References}

\bibliography{ND_Cube.bib}

\end{document}